\documentclass[contrib]{sfibook-gell-mann}
\usepackage{graphicx}  
\usepackage{amssymb}
\tableofcontents
\makeindex
\titlemark{ }
\begin{document}
\newcommand{\EGLOB}{E_{\rm glob}}
\newcommand{\ELOC}{E_{\rm loc}}

\setcounter{page}{1}
\setcounter{tocdepth}{2}
\setcounter{chapter}{0}

\setlength{\hoffset}{0.2in}
\setlength{\textfloatsep}{1.0cm}

\newdimen\captwidth
\captwidth=13cm                           % width captions appear
\newdimen\normalfigwidth
\normalfigwidth=\captwidth                % width figures appear
\newskip\captskip
\captskip=9pt plus3pt minus3pt

\def\normalfigure#1{\hbox to\textwidth{%
        \hfil\resizebox{\normalfigwidth}{!}{\includegraphics{#1}}\hfil}}
\def\capt#1{\refstepcounter{figure}\vskip\captskip\hbox to
\textwidth{%\hfil\vbox{\hsize=\captwidth
\renewcommand{\baselinestretch}{1}{\small
        {\sc Figure \thefigure}\quad#1}\hfil}}

\chapter{The Architecture of Complex Systems}
{Vito Latora$^1$ and Massimo Marchiori$^2$
\\
1) Dipartimento di Fisica e Astronomia, Universit\`a di Catania,
and INFN sezione di Catania, 
%Corso Italia 57, 95129 
Catania, Italy
\\
2) W3C and Lab. for Computer Science,
Massachusetts Institute of Technology, Cambridge, USA}

\chaptermark{The Architecture of Complex Systems}
\markboth{The Architecture of Complex Systems}
{Vito Latora and Massimo Marchiori}

%\begin{abstract}
%Here is the text for an abstract for your chapter.  This section is,
%of course, optional.
%\end{abstract}

%%If your chapter starts with a section rather than text or an
%%abstract, please use the command \firstsection instead of
%%section.

\section{Introduction}
At the present time, the most commonly accepted definition of a
complex system is that of a system containing many interdependent
constituents which interact nonlinearly
\footnote
{The definition may seem
somewhat fuzzy and generic: this is an indication that the
notion of a complex system is still not precisely delineated 
and differs from author to author.
On the other side, there is complete agreement
that the ``ideal'' complex systems are the biological ones,
especially those which have to do with people: our bodies,
social systems, our cultures
\cite{baranger}.
}.
Therefore, when we want to model a complex system,
the first issue has to do with the connectivity
properties of its network, the architecture of the wirings
between the constituents. 
In fact, we have recently learned that the network structure can be
as important as the nonlinear interactions between elements,
and an accurate description of the coupling architecture and
a characterization of the structural properties
of the network can be of fundamental importance
also to understand the dynamics of the system.

In the last few years the research on networks has taken
different directions producing rather unexpected and important
results. Researchers have: 1) proposed various global variables
to describe and characterize the properties of real-world networks; 
2) developed different models to simulate the formation 
and the growth of networks as the ones found in the real world. 
The results obtained can be summed up by saying that statistical 
physics has been able to capture the structure of many diverse 
systems within a few common frameworks, though these
common frameworks are very different from the regular array, or the 
random connectivity, previously used to model the network of a
complex system. 

Here we present a list of some of the global quantities introduced 
to characterize a network: the characteristic path length $L$, the
clustering coefficient $C$, the global efficiency  $\EGLOB$, the
local efficiency $\ELOC$, the cost $Cost$, and the degree
distribution $P(k)$. We also review two classes of networks proposed:   
small-world and scale-free networks. We conclude with a possible 
application of the nonextensive thermodynamics formalism to describe 
scale-free networks.

\section{SMALL-WORLD NETWORKS}
In ref. \cite{watts} Watts and Strogatz have shown that 
the connection topology of some biological, social and 
technological networks is neither completely regular
nor completely random. These networks, 
that are somehow in between regular and random 
networks, have been named {\it small worlds} 
in analogy with the small world phenomenon 
empirically observed in social systems more than 
30 years ago \cite{milgram,newman1}. 
In the mathematical formalism developed by Watts and Strogatz 
a generic network is represented as an unweighted graph $\bf G$
with $N$ nodes (vertices) and $K$ edges (links) between nodes.
Such a graph is described by the adjacency matrix
$\{a_{ij}\}$, whose entry $a_{ij}$ is either $1$ if there is an
edge joining vertex $i$ to vertex $j$, and $0$ otherwise.
The mathematical characterization of the small-world
behavior is based on the evaluation of two quantities,
the characteristic path length $L$ and the clustering
coefficient $C$.

\subsection{The characteristic path length}
The characteristic path length $L$ measures the typical
separation between two generic nodes of a graph $\bf G$.
$L$ is defined as:
$$
L({\bf G})= \frac {1}{N(N-1)} \sum_{i\neq j \in {\bf G} } d_{ij}
\label{l}
$$
where $d_{ij}$ is the shortest path length between $i$ and $j$,
i.e.\ the minimum number of edges traversed to get from a vertex
$i$ to another vertex $j$. By definition $d_{ij} \ge 1$, and
$d_{ij}=1$ if there exists a direct link between $i$ and $j$.
Notice that if $\bf G$ is connected, i.e.\ there exists at least
one path connecting any couple of vertices with a finite number of
steps, then $d_{ij}$ is finite $\forall i \neq j$ and also $L$ is
a finite number. For a non-connected graph, $L$ is an ill-defined quantity,
because it can diverge. This problem is avoided by using 
$\EGLOB$ in place of $L$.

\subsection{The clustering coefficient}
The clustering coefficient $C$ is a local quantity of ${\bf G}$
measuring the average cliquishness of a node. For any node $i$,
the subgraph of neighbors of $i$, $\bf G_i$ is considered. If the
degree of $i$, i.~e.~ the number of edges incident with $i$, is
equal to $k_i$, then $\bf G_i$ is made of $k_i$ nodes and at most
$k_i(k_i-1)/2$ edges. $C_i$ is the fraction of these edges that
actually exist, and $C$ is the average value of $C_i$ all over the
network (by definition $0 \le C \le 1$):
$$
C({\bf G}) = \frac {1}{N} \sum_{i \in {\bf G} } C_i ~~~~~~~~
C_i = {  \mbox{  \# of edges in  $\bf G_i$ }
        \over
        \mbox{  $k_i(k_i-1)/2$ }
     }
$$
The mathematical characterization of the small-world behavior 
proposed by Watts and Strogatz is based on the evaluation of $L$ 
and $C$: small-world networks have high $C$ 
like regular lattices, and short $L$ like random graphs.  
The small-world behavior is ubiquitios in nature and in man-made 
systems. Neural networks, social systems \cite{michelle} 
as the collaboration graph of movie actors \cite{watts} 
or the collaboration network of scientists \cite{newman4}, 
technological networks as the 
World Wide Web or the electrical power grid of the Western US, 
are only few of such examples. 
To give an idea of the numbers obtained we consider the simplest 
case of the neural networks investigated, that of the C.~elegans: 
this network, represented by a graph with $N=282$ nodes (neurons) 
and $K=1974$ edges (connections between neurons), 
gives $L=2.65$ and $C=0.28$ \cite{watts}. 
It is also important to notice that a network as the electrical power grid 
of the western US, can be studied by such a formalism 
only if considered as an unweighted graph, i.e. when no 
importance whatsoever is given to the physical length of 
the links.

\section{EFFICIENT AND ECONOMIC BEHAVIOR}

A more general formalism, valid both for unweighted and weighted
graphs (also non-connected), extends the application of the
small-world analysis to any real complex network, in particular to 
those systems where the euclidian distance between vertices is 
important (as in the case of the electrical power grid of western US), 
and therefore too poorly described only by the topology of connections
\cite{lm2,lm4}. Such systems are better described by two matrices, the
adjacency matrix $\{a_{ij}\}$ defined as before, and a second
matrix $\{\ell_{ij}\}$ containing the weights associated to each
link. The latter is named the  matrix of physical distances, 
because the numbers $\ell_{ij}$ can be imagined as the euclidean 
distances between $i$ and $j$. 
The mathematical characterization 
of the network is based on the evaluation of two quantities, the
global and the local efficiency (replacing $L$ and $C$), and a third 
one quantifying the cost of the network. 
Small worlds are networks that exchange information very 
efficiently both on a global and on a local scale \cite{lm2}.

\subsection{The global efficiency}
In the case of a weighted network the shortest path length
$d_{ij}$ is defined as the smallest sum of the
physical distances throughout all the possible paths in the graph
from $i$ to $j$
\footnote{$\{d_{ij}\}$ is now calculated by using the information
contained both in $\{a_{ij}\}$ and in $\{\ell_{ij}\}$.}.
The efficiency $\epsilon_{ij}$ in the communication between
vertex $i$ and $j$ is assumed to be inversely
proportional to the shortest path length:  $\epsilon_{ij} = 1/d_{ij}$.
When there is no path in the graph
between $i$ and $j$, $d_{ij}=+\infty$ and consistently $\epsilon_{ij}=0$.
Suppose now that every vertex sends information along the network,
through its edges. The global efficiency of
$\bf G$ can be defined as an average of  $\epsilon_{ij}$:
$$
\label{efficiency}
\EGLOB({\bf G})=
\frac{ {{\sum_{{i \ne j\in {\bf G}}}} \epsilon_{ij}}  } {N(N-1)}
          = \frac{1}{N(N-1)}
{\sum_{{i \ne j\in {\bf G}}} \frac{1}{d_{ij}}}
$$
Such a quantity is always a finite number (even when  
${\bf G}$ is unconnected) and 
can be normalized to vary in the range $[0,1]$ if
divided by
$
\EGLOB({\bf G^{ideal}})=
\frac{1}{N(N-1)}
{\sum_{{i \ne j\in {\bf G}}} \frac{1}{l_{ij}}}
$
, the efficiency of the ideal case ${\bf {G^{ideal}}}$ in which 
the graph has all the $N(N-1)/2$ possible edges. In such a case the
information is propagated in the most efficient way since $d_{ij}
=\ell_{ij}~\forall i,j$.

\subsection{The local efficiency}
One of the advantages of the efficiency-based formalism
is that a single measure, the efficiency $E$ (instead of the two
different measures $L$ and $C$) is
sufficient to define the small-world behavior. In fact
the efficiency, can be evaluated for any subgraph of $\bf G$,
in particular for $\bf {G_i}$, the subgraph of the neighbors of $i$
(made by $k_i$ nodes and at most ${k_i(k_i-1)}/2$ edges),
and therefore it can be used also to characterize the local
properties of the graph.
The local efficiency of $\bf G$ is defined as:
$$
\label{localeff}
\ELOC({\bf G})   = \frac{1}{N} \sum_{i \in {\bf G}} ~ {E(\bf {G_i})}
~~~~~~~
{E(\bf {G_i})}
% = \frac{ {{\sum_{{i \ne j\in {\bf G_i}}}} \epsilon^{\prime}_{ij}}  } {k_i(k_i-1)}
          = \frac{1}{k_i(k_i-1)}
{\sum_{{l \ne m\in {{\bf G_i}}}} \frac{1}{d^{\prime}_{lm}}}
$$
where the quantities $\{d^{\prime}_{lm}\}$
are the shortest distances between nodes $l$ and $m$
calculated on the graph $\bf G_i$.
Similarly to $\EGLOB$, also $\ELOC$ can be normalized to
vary in the range $[0,1]$  and plays a role similar to
that of $C$ \cite{lm4}.
Small worlds are networks with high $\EGLOB$ and high $\ELOC$.

\subsection{The Cost}
An important variable to consider, especially when we deal with
weighted networks and when we want to analyze and compare
different real systems, is the cost of a network \cite{lm4}. In
fact, we expect both $\EGLOB$ and $\ELOC$ to be higher ($L$ lower
and $C$ higher) as the number of edges in the graph increases. As
a counterpart, in any real network there is a price to pay for
number and length (weight) of edges. This can be taken into
account by defining the cost of the graph $\bf G$ as the total
length of the network's wirings:
\begin{equation}
Cost({\bf G})= \frac {\sum_{{i \ne j\in {\bf G}}}
a_{ij}~ \ell_{ij}}
                     {\sum_{{i \ne j\in {\bf G}}} { \ell_{ij} }}
\end{equation}
Since the cost of ${\bf G}^{ideal}$ is already included in the
denominator of the formula above,
$Cost$ varies in $[0,1]$ and assumes the maximum
value $1$ when all the edges are present in the graph.
In the case of an unweighted graph, $Cost({\bf G})$ reduces
to the normalized number of edges $2K/N(N-1)$.

With the three variables $\EGLOB$, $\ELOC$ and $Cost$, all defined
in $[0,1]$, it is possible to study in an unified way unweighted
(topological) and weighted networks. And it is possible to define
an economic small world as a network having low $Cost$ and high
$\ELOC$ and $\EGLOB$ (i.e., both economic and small-world). In
figure we report an useful illustrative example obtained by means
of a simple model to construct a class of weighted graphs. We
start by considering a regular network of $N=1000$ nodes placed on
a circle ($\ell_{i,j}$ is given by the euclidean distance between
$i$ and $j$) and $K=1500$ links. A random rewiring procedure is
implemented: it consists in going through each of the links in
turn and independently with some probability $p$ rewire it.
Rewiring means shifting one end of the edge to a new node chosen
randomly with a uniform probability. In this way it is possible to
tune $\bf G$ in a continuous manner from a regular lattice ($p=0$)
into a random graph ($p=1$), without altering the average number
of neighbors equal to $k = 2K/N$. For $p\sim 0.02-0.04$ we observe
the small-world behavior: $\EGLOB$ has almost reached its maximum
value $0.62$ while $\ELOC$ has not changed much from the maximum
value $0.2$ (assumed at $p=0$). Moreover for these values of $p$
the network is also economic, in fact the $Cost$ stays very close
to the minimum possible value (assumed of course in the regular
case $p=0$).
\begin{figure}%Figure 1
\begin{center}
\includegraphics[width=3in,angle=-90]{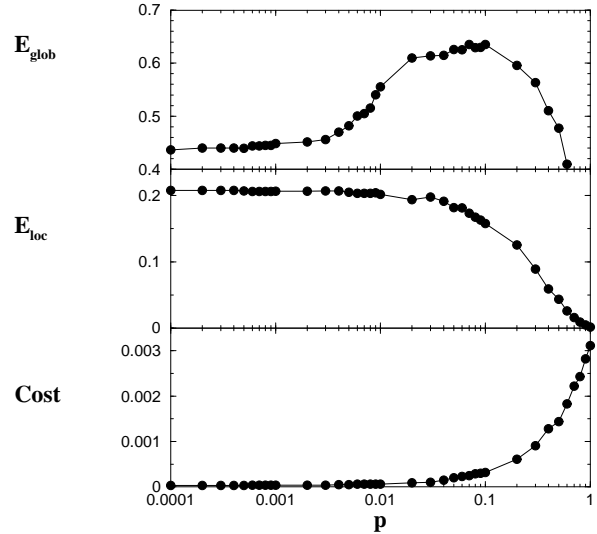}
\caption{The three quantities $\EGLOB$, $\ELOC$ and $Cost$
are reported as function of the rewiring probability $p$ for the
model discussed in the text.
The economic small-world behavior shows up for $p\sim 0.02-0.04$
}
\end{center}
\end{figure}
%
%
%\begin{figure}%Figure 1      confsantafe_apr02_fig1.eps
%\vspace{2in}
%\caption{The three quantities $\EGLOB$, $\ELOC$ and $Cost$
%are reported as function of the rewiring probability $p$ for the
%model discussed in the text.
%The economic small-world behavior shows up for $p\sim 0.02-0.04$
%}
%\end{figure}
%
\\
Some examples of applications to real networks. 
The neural network of the C.~elegans has $\EGLOB=0.35$, 
$\ELOC=0.34$, $Cost = 0.18$: the C.~elegans is an economic 
small world because it achieves high efficiency both at the global 
and local level (about $35\%$ of the global and local efficiency of 
the ideal completely connected case); all of this at a relatively low
cost, with only the $18\%$ of the wirings of the ideal graph. 
As a second example we consider a technological network, the MBTA, 
the Boston underground transportation system. The MBTA is a weighted
network consisting of $N=124$ stations and $K=124$ tunnels 
connecting couples of stations. For such a system we obtain
$\EGLOB=0.63$, $\ELOC=0.03$ and $Cost=0.002$. This means that {\em
MBTA\/} achieves the $63\%$ of the efficiency of the ideal subway
with a cost of only the $0.2\%$. The price to pay for such
low-cost high global efficiency is the lack of local efficiency.
In fact, $\ELOC=0.03$ indicates that, differently from a neural
network (or from a social system), the {\em MBTA\/} is not fault
tolerant, i.e.\ a damage in a station will dramatically affect the
efficiency in the connection between the previous and the next
station. The difference with respect to neural networks comes from
different needs and priorities in the construction and evolution
mechanism. When a subway system is built, the priority is given to
the achievement of global efficiency at a relatively low cost, and
not to fault tolerance. In fact a temporary problem in a station
can be solved in an economic way by other means: for example,
walking, or taking a bus from the previous to the next station.
%Changing the {\em MBTA\/} network to take into account, for
%example the bus system, this extended transportation system comes
%back to be an economic small-world network. In fact this extended
%transportation system achieves high global but also high local
%efficiency ($\EGLOB=0.72$, $\ELOC=0.43$), at a still low price
%($Cost$ has only increased from $0.002$ to $0.004$). This can be
%explained by the fact that in such a system we consider almost all
%the reasonable transportation alternatives available at that
%scale. In this way the system is closed, there are no other
%reasonable routing alternatives, and so fault-tolerance comes
%back, after the cost, as a leading construction principle.
Applications to other real networks can be found in
ref.\cite{lm4}.

\section{SCALE-FREE NETWORKS}

\subsection{Degree Distribution}
Other important information on a network 
can be extracted from its degree distribution $P(k)$. 
The latter is defined as the probability of
finding nodes with $k$ links: $P(k)=\frac{N(k)}{N}$, where $N(k)$
is the number of nodes with $k$ links. 
Many large networks, as the World Wide Web, the Internet, 
metabolic and protein networks have been 
named {\it scale-free} networks because their degree
distribution follows a power-law for large $k$
\cite{barabasi2,barabasi4met}. Also a social system of interest for the
spreading of sexually transmitted diseases \cite{sta}, and the
connectivity network of atomic clusters' systems \cite{doye} 
show a similar behavior. The most interesting fact is that
neither regular nor random graphs display long tails in P(k), and 
the presence of nodes with large $k$ strongly affects 
the properties of the network \cite{ves}, as for instance its 
response to external factors \cite{crucitti}. 
In ref.\cite{barabasi2} Barabasi and Albert have proposed a simple
model (the BA model) to reproduce the P(k) found in
real networks by modelling the dynamical growth of the network.
The model is based on two simple mechanisms, growth and
preferential attachment, that are also the main ingredients
present in the dynamical evolution of the real-world networks. As
an example, the World Wide Web grows in time by the
addition of new web pages, and a new web page will more
likely include hyperlinks to popular documents with already high
degree. 
%, because such highly connected documents are easy to find
%and thus well known. 
Starting by an initial network with a few nodes and adding new nodes 
with new links preferentially connected
to the most important existing nodes, the dynamics of the BA model
produces (in the stationary regime) scale-free networks with a
power-law degree distribution \cite{barabasi2}:
$$
P(k) \sim k^{-\gamma} ~~~\gamma=3
$$
The model predicts the emergence of the scale-free behavior 
observed in real networks, though the exponents in the power law 
of real networks can be different from 3 
(usually it is in the range between 2 and 3).

\subsection{Nonextensive Statistical Mechanics} 
A more careful analysis of the shape of P(k) of 
many of the real networks considered evidentiates the presence of a 
plateau for small k. See for example fig.1a of 
Ref.\cite{barabasi2} and fig.2b of Ref. \cite{sta}. We have
observed that such a plateau for small $k$ and the different slopes 
of the power-law for large $k$ can be perfectly reproduced by
using the generalized power-law distribution 
$$
P(k) \sim [1+(q-1)\beta k]^{\frac{1}{1-q}}
$$
with two fitting parameters: $q$ related to the slope of the
power law for large $k$, and $\beta$ \cite{next}. 
The generalized probability distribution above can be obtained as a
stationary solution of a generalized Fokker-Planck equation
with a nonlinear diffusion term \cite{plastino}. 
We therefore believe that is possible to rephrase the generalized
Fokker-Planck equation in terms of a generalized mechanism of
network construction, and to implement
a model (more general than the BA model) able to reproduce the
plateau and the different slopes
of P(k).

\section*{ACKNOWLEDGMENTS}
We thank E. Borges, P. Crucitti, M.E.J. Newman, A. Rapisarda, 
C. Tsallis and G. West for their useful comments.

%\clearpage

\end{document}